\documentclass[11pt]{article}
\usepackage{epsfig} 
\newcommand{\Dslash}{D \! \! \! \! /}

\newcommand{\half}{\mbox{\small{$\frac{1}{2}$}}} 
 
\newcommand{\Nf}{N_{\!f}} 
\newcommand{\NA}{N_{\!A}} 
\newcommand{\NF}{N_{\!F}} 
\newcommand{\MSbar}{\overline{\mbox{MS}}} 

\begin{document}
\title{Two loop correction to the Gribov mass gap equation in Landau gauge QCD} 
\author{J.A. Gracey, \\ Theoretical Physics Division, \\ 
Department of Mathematical Sciences, \\ University of Liverpool, \\ P.O. Box 
147, \\ Liverpool, \\ L69 3BX, \\ United Kingdom.} 
\date{} 
\maketitle 
\vspace{5cm} 
\noindent 
{\bf Abstract.} We determine the two loop correction to Gribov's mass gap 
equation for quantum chromodynamics in the Landau gauge in the $\MSbar$ scheme 
by computing the two loop correction to the horizon condition derived from 
Zwanziger's local renormalizable Lagrangian which incorporates the Gribov 
parameter. We verify that with the explicit result, the two loop ghost
propagator is enchanced in the infrared. 

\vspace{-17.5cm}
\hspace{13.5cm}
{\bf LTH 677}

\newpage 

In 1978 Gribov demonstrated that globally fixing a gauge in a non-abelian
gauge theory could not be performed in a unique way, \cite{1}. This is due to 
the fact that different gauge configurations could satisfy the same gauge 
fixing condition. Related to this was the observation that the first zero mode 
of the Faddeev-Popov operator defined the edge or horizon of a region, known as
the Gribov volume, inside which one had to restrict the region of integration 
of the path integral defining the quantum field theory, \cite{1}. One 
consequence of the existence of this restricted domain was that the form of the
gluon and ghost propagators in the infrared regime differed significantly from 
that usually used in perturbative, and therefore ultraviolet, calculations. 
More concretely, the ghost propagator was enhanced and the gluon propagator 
suppressed as the momentum decreases. Such behaviour is believed to be an 
element of the confinement mechanism which is nonperturbative. Another feature 
of Gribov's original work was the definition of the volume of the Gribov 
region\footnote{For the reader interested in more detailed background to the
Gribov problem, the lectures of \cite{2} give detailed working of the
calculations of \cite{1}}. This was quantified in terms of a dimensionful 
parameter, known as the Gribov volume or mass, which was not independent but 
satisfied a gap equation. Gribov computed the explicit form of the gap 
equation to one loop and related it to the running coupling constant, 
\cite{1}. Subsequently, Zwanziger examined the Gribov path integral 
formulation of the problem and managed to construct a localized renormalizable 
Lagrangian, which included the Gribov mass explicitly, in addition to several
new ghost fields over and above the usual Faddeev-Popov ghosts originating 
from the standard gauge fixing procedure, \cite{3,4,5,6}. When these new 
ghosts are eliminated by their equations of motion one recovers the non-local 
Lagrangian introduced by Gribov. The renormalizability properties of the 
Zwanziger Lagrangian have been studied in detail in \cite{7,8} in the Landau 
gauge and there is an interesting structure. First, the wave function 
renormalization constants of the extra Zwanziger ghosts are identical to those 
of the Faddeev-Popov ghosts in the Landau gauge despite being spin-$1$ fields
with a different colour structure. Second, the renormalization of the Gribov 
mass is not independent. Specifically the anomalous dimension of the Gribov 
mass parameter is given by a combination of the $\beta$-function and gluon 
anomalous dimension. These have been demonstrated to all orders in 
perturbation theory using the algebraic renormalization technique, \cite{7,8}. 
That Gribov's original formulation of the limitations of the quantization of a 
non-abelian gauge theory can be recast in a localized renormalizable 
Lagrangian means that it ought to be possible to use the Zwanziger Lagrangian, 
\cite{3,4,5,6}, to study problems in the presence of the Gribov mass. For 
example, it should be possible as an initial exercise in this direction to 
compute the {\em two} loop correction to Gribov's mass gap equation and
therefore refine the relation between the Gribov mass and the coupling 
constant. This is the main aim of this article where we will concentrate on 
establishing the two loop correction to Gribov's mass gap equation in the 
Landau gauge in quantum chromodynamics, QCD.  

First, we recall the Zwanziger Lagrangian in the conventions and notation we 
will use. For QCD we have in $d$ spacetime dimensions, \cite{3,4,5,6}, 
\begin{eqnarray}
L^Z &=& L^{\mbox{\footnotesize{QCD}}} ~+~ \bar{\phi}^{ab \, \mu} \partial^\nu
\left( D_\nu \phi_\mu \right)^{ab} ~-~ \bar{\omega}^{ab \, \mu} \partial^\nu 
\left( D_\nu \omega_\mu \right)^{ab} \nonumber \\  
&& -~ g f^{abc} \partial^\nu \bar{\omega}^{ae} \left( D_\nu c \right)^b
\phi^{ec \, \mu} ~-~ \frac{\gamma^2}{\sqrt{2}} \left( f^{abc} A^{a \, \mu} 
\phi^{bc}_\mu ~+~ f^{abc} A^{a \, \mu} \bar{\phi}^{bc}_\mu \right) ~-~ 
\frac{d \NA \gamma^4}{2g^2} 
\label{lagz} 
\end{eqnarray} 
where $\gamma$ is the Gribov mass parameter and  
\begin{equation} 
L^{\mbox{\footnotesize{QCD}}} ~=~ -~ \frac{1}{4} G_{\mu\nu}^a 
G^{a \, \mu\nu} ~-~ \frac{1}{2\alpha} (\partial^\mu A^a_\mu)^2 ~-~ 
\bar{c}^a \partial^\mu D_\mu c^a ~+~ i \bar{\psi}^{iI} \Dslash \psi^{iI} 
\end{equation} 
where $A^a_\mu$ is the gauge field, $c^a$ and $\bar{c}^a$ are the Faddeev-Popov
ghosts and it is understood that we will only consider the Landau gauge given 
by $\alpha$~$=$~$0$. Further, the Zwanziger ghost fields, $\{\phi^{ab}_\mu, 
\bar{\phi}^{ab}_\mu\}$, are complex conjugate commuting fields and 
$\{ \omega^{ab}_\mu, \bar{\omega}^{ab}_\mu \}$ are complex conjugate 
anti-commuting fields. The remaining conventions are that 
$G^a_{\mu\nu}$~$=$~$\partial_\mu A^a_\nu$~$-$~$\partial_\nu A^a_\nu$~$-$
$g f^{abc} A^b_\mu A^c_\nu$, $g$ is the coupling constant, $T^a$ are the
colour group generators whose structure constants are $f^{abc}$, $\psi^{iI}$ 
is the (massless) quark field, $1$~$\leq$~$a$~$\leq$~$\NA$, 
$1$~$\leq$~$I$~$\leq$~$\NF$ and $1$~$\leq$~$i$~$\leq$~$\Nf$ with $\NF$ and 
$\NA$ the dimensions of the fundamental and adjoint representations 
respectively, and $\Nf$ is the number of quark flavours. The covariant 
derivatives are defined by  
\begin{eqnarray} 
D_\mu c^a &=& \partial_\mu c^a ~-~ g f^{abc} A^b_\mu c^c ~~,~~ 
D_\mu \psi^{iI} ~=~ \partial_\mu \psi^{iI} ~+~ i g T^a A^a_\mu \psi^{iI} 
\nonumber \\
\left( D_\mu \phi_\nu \right)^{ab} &=& \partial_\mu \phi^{ab}_\nu ~-~
g f^{acd} A^c_\mu \phi^{db}_\nu ~. 
\end{eqnarray} 
It is worth noting the notational contrasts between this version of the 
Lagrangian and that of other articles. We have defined the Gribov mass 
parameter, $\gamma$, to be of mass dimension one. Some authors use $\gamma^2$ 
or $\gamma^4$, in our notation, as the equivalent Gribov parameter with 
respective mass dimension two and four. Second, we do not include a coupling 
constant with $\gamma$ in the mixed mass term for the same reasons one does 
not include it for, say, a quark mass term. Third, the appearance of the
factor $\frac{1}{\sqrt{2}}$ in the mixed quadratic term plays a key role not 
only in the derivation of the propagators of (\ref{lagz}) but in verifying the
correctness of the gap equation we will determine. With this version of the 
Gribov-Zwanziger Lagrangian, we follow \cite{6} in noting that the Gribov 
horizon condition, which leads to the mass gap, is 
\begin{equation}
f^{abc} \langle A^{a \, \mu}(x) \phi^{bc}_\mu(x) \rangle ~=~ 
\frac{d \NA \gamma^2}{\sqrt{2}g^2} 
\label{hordef} 
\end{equation} 
where all quantities are bare and (\ref{hordef}) is derived from the equation 
of motion of $\bar{\phi}_\mu^{ab}$ in (\ref{lagz}). 

At one loop it is elementary to see that this condition leads to Gribov's
original gap equation of \cite{1}. For instance, from the terms quadratic in 
the fields of (\ref{lagz}), the gluon and commuting ghost propagators are 
given by 
\begin{eqnarray}
\langle A^a_\mu(p) A^b_\nu(-p) \rangle &=& -~ 
\frac{\delta^{ab}p^2}{[(p^2)^2+C_A\gamma^4]} P_{\mu\nu}(p) \nonumber \\  
\langle A^a_\mu(p) \bar{\phi}^{bc}_\nu(-p) \rangle &=& -~ 
\frac{f^{abc}\gamma^2}{\sqrt{2}[(p^2)^2+C_A\gamma^4]} P_{\mu\nu}(p) 
\nonumber \\  
\langle \phi^{ab}_\mu(p) \bar{\phi}^{cd}_\nu(-p) \rangle &=& -~ 
\frac{\delta^{ac}\delta^{bd}}{p^2}\eta_{\mu\nu} ~+~  
\frac{f^{abe}f^{cde}\gamma^4}{p^2[(p^2)^2+C_A\gamma^4]} P_{\mu\nu}(p) 
\nonumber \\ 
\langle \omega^{ab}_\mu(p) \bar{\omega}^{cd}_\nu(-p) \rangle &=& -~ 
\frac{\delta^{ac}\delta^{bd}}{p^2} \eta_{\mu\nu} 
\label{mixprop} 
\end{eqnarray} 
in momentum space where
\begin{equation}
P_{\mu\nu}(p) ~=~ \eta_{\mu\nu} ~-~ \frac{p_\mu p_\nu}{p^2} 
\end{equation}
and $f^{acd} f^{bcd}$ $=$ $C_A \delta^{ab}$. Concerning conventions here the 
appearance of $C_A$ in the denominator factors is a direct consequence of our 
choice of the coefficient of the mixed quadratic term of (\ref{lagz}). In 
\cite{4} the convention was to define the mixed quadratic term to have an 
additional factor proportional to $\sqrt{C_A}$ whence the corresponding term of
(\ref{mixprop}) would be merely $\gamma^4$. However, the key point in the 
derivation of (\ref{mixprop}) is the role played by the factor of 
$\frac{1}{\sqrt{2}}$ in the mixed mass term of (\ref{lagz}). Ordinarily in 
deriving the propagators from a Lagrangian one isolates the terms quadratic in 
the fields in momentum space and inverts the associated operator or matrix. In 
the case of (\ref{lagz}) this procedure is followed but not only is one dealing
with a mixing between the gluon and commuting ghost fields but the former is 
real whilst the latter is complex. To deal with this consistently, one must 
write this matrix with respect to the basis $(\frac{1}{\sqrt{2}} A^a_\mu, 
\phi^{ab}_\mu)$. In the absence of the mixing term one would not need to do 
this as the matrix to be inverted is block diagonal. If, by contrast, the 
factor of $\frac{1}{\sqrt{2}}$ was omitted from (\ref{lagz}), then the common 
factor in the propagators of (\ref{mixprop}) would be $[(p^2)^2 + 2 C_A 
\gamma^4]$ as was used in \cite{8}. Since the original Gribov article has the 
factor $[(p^2)^2 + C_A \gamma^4]$ we choose to include the normalization of 
$\frac{1}{\sqrt{2}}$ explicitly in (\ref{lagz}) to have propagators which are 
consistent with those of \cite{1}. We will comment further on the significance 
of this factor later, except to note here that the appearance of 
$\frac{1}{\sqrt{2}}$ in (\ref{mixprop}) derives from the form of the basis
chosen. Finally, concerning the mixed propagator of (\ref{mixprop}), we note 
that this mixing disappears in either the limit as $\gamma$~$\rightarrow$~$0$ 
or in the abelian limit which is formally defined as 
$f^{abc}$~$\rightarrow$~$0$ implying $C_A$~$\rightarrow$~$0$.

Whilst this type of mixed propagator will complicate the problem of performing 
loop calculations with (\ref{lagz}), at one loop it is straightforward to use 
the mixed $A^a_\mu$-$\phi^{bc}_\nu$ propagator itself to evaluate the vacuum 
expectation value of (\ref{hordef}). Thoughout our loop calculations we will 
use dimensional regularization in $d$~$=$~$4$~$-$~$2\epsilon$ dimensions and 
subtract the divergences using the $\MSbar$ scheme. Using partial fractions 
and the standard one loop massive integral   
\begin{equation}
\int_k \frac{1}{[k^2+m^2]} ~=~ \frac{(m^2)^{\half d - 1}}{(4\pi)^{\half d}}
\Gamma \left( 1-\half d \right) 
\label{int1} 
\end{equation}
where $\int_k$ $=$ $\int \frac{d^dk}{(2\pi)^d}$ and $m$ is a general mass
argument, we reproduce the finite Gribov mass gap equation in four dimensions
of \cite{1} as  
\begin{equation} 
1 ~=~ C_A \left[ \frac{5}{8} - \frac{3}{8} \ln \left( 
\frac{C_A\gamma^4}{\mu^4} \right) \right] a ~+~ O(a^2) 
\label{grib1} 
\end{equation} 
where $a$ $=$ $g^2/(16\pi^2)$ and $\mu$ is the $\MSbar$ renormalization scale 
introduced to retain a dimensionless coupling constant in $d$-dimensions and 
incorporates the usual numerical factor of 
$4\pi e^{-\gamma_{\mbox{\footnotesize{E}}}}$ with 
$\gamma_{\mbox{\footnotesize{E}}}$ the Euler-Mascheroni constant. In deriving 
this finite expression we have introduced renormalization constants for the 
bare objects appearing in (\ref{hordef}). Specifically these are defined by 
\begin{eqnarray} 
A^{a \, \mu}_{\mbox{\footnotesize{o}}} &=& \sqrt{Z_A} \, A^{a \, \mu} ~~,~~ 
c^a_{\mbox{\footnotesize{o}}} ~=~ \sqrt{Z_c} \, c^a ~~,~~ 
\phi^{ab}_{\mu \, \mbox{\footnotesize{o}}} ~=~ \sqrt{Z_\phi} \, 
\phi^{ab}_\mu ~~,~~ 
\omega^{ab}_{\mu \, \mbox{\footnotesize{o}}} ~=~ \sqrt{Z_\omega} \, 
\omega^{ab}_\mu \nonumber \\ 
\psi_{\mbox{\footnotesize{o}}} &=& \sqrt{Z_\psi} \psi ~~,~~ 
g_{\mbox{\footnotesize{o}}} ~=~ Z_g \, g ~~,~~ 
\gamma_{\mbox{\footnotesize{o}}} ~=~ Z_\gamma \, \gamma 
\label{rencon} 
\end{eqnarray} 
where the subscript ${}_{\mbox{\footnotesize{o}}}$ denotes the bare quantity
here. Since the main aim of this article is to extend (\ref{grib1}) to the 
next order, we have made use of the symbolic manipulation language {\sc Form}, 
\cite{9}, to first reproduce the original mass gap of \cite{1} before 
considering the two loop calculation. In order to check that we have used a 
consistent set of Feynman rules in the {\sc Form} programme, we have first 
explicitly renormalized (\ref{lagz}) to two loops in $\MSbar$ and verified 
that the Slavnov-Taylor identities derived in \cite{7,8} are correctly 
reproduced. However, such a two loop renormalization can be performed with the
massless version of (\ref{lagz}). This means that the {\sc Mincer}
algorithm for massless $2$-point functions, \cite{10}, was the tool used for
this particular computation. For completeness we note that the explicit 
expressions for the renormalization constants for the quantities which do not 
ordinarily arise in the treatment of the usual QCD Lagrangian are  
\begin{eqnarray} 
Z_\phi ~=~ Z_\omega &=& 1 ~+~ \frac{3C_A}{4} \frac{a}{\epsilon} \nonumber \\
&& +~ \left[ \left( \frac{1}{2} C_A T_F \Nf - \frac{35}{32} C_A^2 \right) 
\frac{1}{\epsilon^2} ~+~ \left( \frac{95}{96} C_A^2 - \frac{5}{12} 
C_A T_F \Nf \right) \frac{1}{\epsilon} \right] a^2 ~+~ O(a^3) \nonumber \\ 
\end{eqnarray} 
and 
\begin{eqnarray} 
Z_\gamma &=& 1 ~+~ \left[ \frac{1}{3} T_F \Nf  - \frac{35}{48} C_A \right] 
\frac{a}{\epsilon} \nonumber \\
&& +~ \left[ \left( \frac{7385}{4608} C_A^2 + \frac{5}{18} T_F^2 \Nf^2 ~-~ 
\frac{193}{144} C_A T_F \Nf \right) \frac{1}{\epsilon^2} \right.  \nonumber \\
&& \left. ~~~~~+~ \left( C_F T_F \Nf + \frac{35}{48} C_A T_F \Nf - 
\frac{449}{384} C_A^2 \right) \frac{1}{\epsilon} \right] a^2 ~+~ O(a^3) 
\end{eqnarray} 
where $T^a T^a$ $=$ $C_F I$ and $\mbox{tr} \left( T^a T^b \right)$ $=$ $T_F
\delta^{ab}$. The pole structure can be encoded in the renormalization group 
functions as 
\begin{eqnarray} 
\gamma_\phi(a) &=& \gamma_\omega(a) ~=~  -~ \frac{3}{4} C_A a ~+~ \left[ 
40 C_A T_F \Nf - 95 C_A^2 \right] \frac{a^2}{48} ~+~ O(a^3) \nonumber \\  
\gamma_\gamma(a) &=& \left[ 16 T_F \Nf - 35 C_A \right] \frac{a}{48} ~+~ 
\left[ 280 C_A T_F \Nf - 449 C_A^2 + 192 C_F T_F \Nf \right]
\frac{a^2}{192} ~+~ O(a^3) \nonumber \\  
\end{eqnarray} 
where, in the Landau gauge,  
\begin{equation}
\gamma_\phi(a) ~=~ \beta(a) \frac{\partial \ln Z_\phi}{\partial a} ~~,~~  
\gamma_\omega(a) ~=~ \beta(a) \frac{\partial \ln Z_\omega}{\partial a} ~~,~~  
\gamma_\gamma(a) ~=~ \mu \frac{\partial \ln \gamma}{\partial \mu} 
\label{gamdef}
\end{equation} 
and in (\ref{gamdef}) we have used the fact that the renormalization constants 
do not depend on $\gamma$. Clearly $\gamma_\phi(a)$ and $\gamma_\omega(a)$ 
are equivalent to $\gamma_c(a)$ in agreement with the expectation of 
\cite{7,8} and $\gamma_\gamma(a)$ satisfies the Slavnov-Taylor identity
derived in \cite{7,8}.

Concerning the explicit expression (\ref{grib1}) we note that it agrees with
that of \cite{1} when evaluated in dimensional regularization in $\MSbar$. In 
this respect it is important to note the role played by the explicit appearance
of the dimension $d$ in (\ref{grib1}). In evaluating (\ref{hordef}) we have not
set $d$ to be $4$ on the right hand side at the outset. Whilst this may appear 
to be a trivial point here, since the contribution from the $O(\epsilon)$ term 
of $d$~$=$~$4$~$-$~$2\epsilon$ is present in the constant term of 
(\ref{grib1}), it will turn out that to have a {\em finite} {\em two} loop gap 
equation one must retain $d$ as $4$~$-$~$2\epsilon$ in (\ref{hordef}) prior to 
renormalization in $\MSbar$. Having established (\ref{grib1}) it is worth 
noting that for an abelian theory the equation is trivially satisfied since 
$\NA$ is formally zero in the original defining horizon condition, 
(\ref{hordef}). 
 
Zwanziger's elegant reformulation of the Gribov problem and the horizon 
definition, (\ref{hordef}), in his Lagrangian immediately opens up the path to 
computing the two loop correction to (\ref{grib1}) which we now detail. One 
simply has to evaluate the two loop corrections to the vacuum expectation 
value $f^{abc} \langle A^{a \, \mu} \phi^{bc}_\mu \rangle$. Since this will 
involve two loop massive vacuum bubbles it remains to determine the set of 
Feynman diagrams and the procedure to evaluate them. One problem with the 
former is the complication of having a mixed $A^a_\mu$-$\phi^{bc}_\nu$ 
propagator. To correctly establish the Feynman graphs we used the {\sc Qgraf} 
package, \cite{11}, using an adaptation provided by the author\footnote{We are 
very much indebted to Dr P. Nogueira for his elegant solution of the mixed 
propagator problem in {\sc Qgraf}.} of \cite{11}. Consequently there are $17$ 
two loop Feynman diagrams to determine. These were computed using a symbolic 
manipulation programme written in {\sc Form} which took the {\sc Qgraf} 
output, substituted the Feynman rules for (\ref{lagz}) and broke up the vacuum 
bubbles into a common form, which was  
\begin{equation}
I_2\left( m_1^2, m_2^2, m_3^2; a,b,c \right) ~=~ \int_{kl} \frac{1}
{[k^2+m_1^2]^a [l^2+m_2^2]^b [(k-l)^2+m_3^2]^c} 
\label{int2} 
\end{equation} 
in addition to the product of two one loop integrals of the form of
(\ref{int1}). Here $\{a,b,c\}$ are strictly positive integers and the two loop 
vacuum bubbles have potentially three mass scales after one applies partial 
fractioning. In other words the set of masses $\{ m_1^2, m_2^2, m_3^2 \}$ can 
take any combination of values in the set $\{ 0, i \gamma^2, - i \gamma^2 \}$. 
Since the general form of (\ref{int2}) has been evaluated to the finite part in
$d$-dimensions for arbitrary $m_1$, $m_2$ and $m_3$, it was a straightforward
exercise to make the appropriate identifications in a {\sc Form} routine 
using the results of \cite{12,13,14}. Though it is worth noting that whilst 
the combinations of $\{ m_1^2, m_2^2, m_3^2 \}$ may appear to lead to a 
complex value for the integrals, in the overall sum for $f^{abc} 
\langle A^{a \, \mu} \phi^{bc}_\mu \rangle$ the final result remained real 
which was a useful check. Further checks resided in the fact that the correct 
divergence structure resulting from combining the basic integrals (\ref{int1}) 
and (\ref{int2}) in summing up all the contributions from all the Feynman 
diagrams was correctly cancelled by the available renormalization constants. 
In this respect we have followed the standard procedure of \cite{15} for 
renormalization in automatic calculations. In this the two loop calculation is 
performed for bare parameters before their renormalized versions are 
introduced which automatically introduce the appropriate counterterms. 
Consequently, one is left with the main result of this article which is the 
finite two loop correction to (\ref{grib1}), 
\begin{eqnarray} 
1 &=& C_A \left[ \frac{5}{8} - \frac{3}{8} \ln \left( 
\frac{C_A\gamma^4}{\mu^4} \right) \right] a \nonumber \\ 
&& +~ \left[ C_A^2 \left( \frac{2017}{768} - \frac{11097}{2048} s_2
+ \frac{95}{256} \zeta(2)
- \frac{65}{48} \ln \left( \frac{C_A\gamma^4}{\mu^4} \right)
+ \frac{35}{128} \left( \ln \left( \frac{C_A\gamma^4}{\mu^4} \right)
\right)^2 \right. \right. \nonumber \\
&& \left. \left. ~~~~~~~~~~~~+~ \frac{1137}{2560} \sqrt{5} \zeta(2) 
- \frac{205\pi^2}{512} \right) \right. \nonumber \\
&& \left. ~~~~~+~ C_A T_F \Nf \left( -~ \frac{25}{24} - \zeta(2)
+ \frac{7}{12} \ln \left( \frac{C_A\gamma^4}{\mu^4} \right)
- \frac{1}{8} \left( \ln \left( \frac{C_A\gamma^4}{\mu^4} \right) \right)^2 
+ \frac{\pi^2}{8} \right) \right] a^2 \nonumber \\
&& +~ O(a^3)  
\label{grib2} 
\end{eqnarray} 
where $s_2$ $=$ $(2\sqrt{3}/9) \mbox{Cl}_2(2\pi/3)$, $\mbox{Cl}_2(x)$ is the
Clausen function and $\zeta(n)$ is the Riemann zeta function. For the 
interested reader $s_2$ and $\sqrt{5} \zeta(2)$ arise from the finite part of 
integrals such as $I_2(m^2,m^2,m^2;1,1,1)$ and $I_2(m^2,m^2,-m^2;1,1,1)$
respectively. We also note that taking $d$ as $4$ in (\ref{hordef}) would have
resulted in an extra divergence which could not be cancelled.  

Having established the two loop gap equation for the Gribov mass, we now 
comment on its implication for the ghost propagator. In \cite{1} Gribov
showed that the one loop condition (\ref{grib1}) ensured that the one loop
correction to the ghost propagator was enhanced in the infrared. More
specifically if one writes the full ghost propagator as
\begin{equation}
G^{ab}_c(p^2) ~=~ \frac{\delta^{ab}}{p^2[1+u(p^2)]}
\end{equation}
then in the ultraviolet, the ghost propagator has the usual (perturbative)
behaviour of $1/p^2$. However, in the infrared the propagator behaves as
$1/(p^2)^2$ as $p^2$ $\rightarrow$ $0$, \cite{1}. Gribov indicated that this
was a property of a confining non-abelian gauge theory. Moreover, he computed
the one loop correction to the ghost propagator to $O(p^2)$ and found that
$1$~$+$~$u(0)$~$=$~$0$ provided that $\gamma$ satisfied (\ref{grib1}). This 
observation of $1$~$+$~$u(0)$~$=$~$0$, which is known as the Kugo-Ojima
confinement condition, \cite{16}, should also hold at two loop in (\ref{lagz})
if the Gribov-Zwanziger Lagrangian is a consistent formulation of the Gribov
horizon condition. Therefore, we have computed the two loop correction to
$u(p^2)$ by generating the $31$ Feynman diagrams using {\sc Qgraf}, expanding
them to $O(p^2)$ and evaluating the resulting vacuum bubbles. It transpires 
that this $O(p^2)$ correction to $u(p^2)$ is {\em exactly} the same as the two
loop part of (\ref{grib2}) and therefore the Kugo-Ojima condition is satisfied
at two loops {\em precisely} for all colour groups and $\Nf$ massless quarks. 
Hence, the ghost propagator is enhanced at two loops in the Gribov-Zwanziger 
Lagrangian and has a $1/(p^2)^2$ behaviour in the infrared which is in 
qualitative agreement with other approaches. For instance, ghost enhancement
has also been observed in explicit (non-perturbative) studies using the 
Schwinger-Dyson formalism and on the lattice. More specifically in the 
Schwinger-Dyson computations estimates for the exponent of the power-law 
behaviour of the gluon and ghost propagators in the infrared have been 
extracted and give values which are different from those of the ultraviolet 
form of the propagator, \cite{17,18}. Also, in several lattice studies the 
Kugo-Ojima confinement condition itself has also been examined. See, for 
instance, \cite{19,20,21,22}, where one recent lattice estimate for $u(0)$ is 
$-$~$0.83$, \cite{22}. 

It is worth commenting on one final aspect of the two loop calculation of 
(\ref{grib2}). If one uses the propagators of (\ref{lagz}) without the factor 
of $\frac{1}{\sqrt{2}}$ in (\ref{mixprop}) then not only would the ghost 
propagator not be enhanced but at an earlier point of the study, the ghost 
$2$-point function would not actually be {\em finite} either. This further 
justifies the derivation of (\ref{mixprop}) from (\ref{lagz}) and what would 
maybe initially appear as a peculiar convention in the definition of the mass 
term. More importantly, that (\ref{grib2}) is exactly what is required for 
ghost enhancement provides a strong check on the result (\ref{grib2}) as well 
as ensuring that one has a consistent set of Feynman rules for non-zero
$\gamma$. Next, in concentrating on the consistency aspect of the ghost with 
respect to enhancement, the gluon propagator is also expected to have infrared 
behaviour different from the usual ultraviolet form, \cite{1,18}. Whilst we 
have not examined the two loop corrections to the gluon propagator as it is 
technically more difficult to analyse than the ghost in the 
$p^2$~$\rightarrow$~$0$ limit, we do not believe its behaviour in the 
Gribov-Zwanziger Lagrangian, (\ref{lagz}), to be inconsistent with the 
expectation that the propagator vanishes, \cite{1,18}. 

In conclusion, we have provided the $O(a^2)$ correction to the Gribov mass
gap equation for QCD in the Landau gauge. Reassuringly the explicit form,
(\ref{grib2}), guarantees the enhancement of the ghost propagator at two loops.
Given the recent interest in both the Zwanziger approach to incorporating the 
Gribov problem in a localized renormalizable Lagrangian and in gauges such as 
linear covariant, \cite{23}, and maximal abelian gauge, \cite{24}, it would 
appear plausible that one could extend the one loop mass gap equations in 
those gauges to two loops as well. Moreover, given that \cite{8,25,26} also 
examined the Gribov problem using the local composite operator formalism to 
include the dimension two composite operator $\half A^{a \, \mu} A^a_\mu$ it 
would be interesting to extend that one loop analysis to see whether the 
operator condenses and lowers the vacuum energy as it does in the case when 
the Gribov volume is regarded as infinite. Finally, we note that we believe 
that this is the first non-trivial loop computation performed with Zwanziger's 
Lagrangian. Given that the Gribov mass can be incorporated in calculations 
now, it would be interesting to examine the effects the presence of $\gamma$ 
has in phenomenological analyses. For instance, using QCD to examine deep 
inelastic scattering problems one will inevitably wish to extend such analyses 
towards the infrared. It is important that the Gribov limitations are taken 
into account because theoretical predictions may no longer be relevant. One 
case in point is the extraction of estimates for the dimension four condensate 
$\langle G_{\mu\nu}^a G^{a \, \mu\nu} \rangle$. Since the expansion of the 
gluon propagator in the presence of the Gribov mass naturally gives rise to a 
power correction of dimension four, it would seem important that the 
consequences of a non-zero $\gamma$ are understood in the corresponding 
underlying operator product expansion. 

\vspace{0.3cm} 
\noindent
{\bf Acknowledgement.} The author thanks Prof. S. Sorella and Dr D. Dudal for
discussions concerning the Gribov problem. Also the author is grateful to Dr P.
Nogueira for solving the problem of mixed propagators in the {\sc Qgraf}
package \cite{11} and the organisers of the ACAT05 workshop where the latter
discussions took place.

\end{document}